\documentclass[aps,prb,twocolumn,floats,showpacs,superscriptaddress]{revtex4}
\usepackage{graphicx,epsfig}
\usepackage{times}
\usepackage{graphics,dcolumn,bm,float}
\usepackage{amssymb,amsmath,rotate,color}

\begin{document}
\unitlength 1 cm
\newcommand{\be}{\begin{equation}}
\newcommand{\ee}{\end{equation}}
\newcommand{\bearr}{\begin{eqnarray}}
\newcommand{\eearr}{\end{eqnarray}}
\newcommand{\nn}{\nonumber}
\newcommand{\vk}{\vec k}
\newcommand{\vp}{\vec p}
\newcommand{\vq}{\vec q}
\newcommand{\vkp}{\vec {k'}}
\newcommand{\vpp}{\vec {p'}}
\newcommand{\vqp}{\vec {q'}}
\newcommand{\bk}{{\bf k}}
\newcommand{\bp}{{\bf p}}
\newcommand{\bq}{{\bf q}}
\newcommand{\br}{{\bf r}}
\newcommand{\bR}{{\bf R}}
\newcommand{\up}{\uparrow}
\newcommand{\down}{\downarrow}
\newcommand{\fns}{\footnotesize}
\newcommand{\ns}{\normalsize}
\newcommand{\cdag}{c^{\dagger}}

\title{
Excitation Spectrum of One-dimensional Extended Ionic Hubbard Model}

\author{M. Hafez{\footnote{Electronic address: m.hafeztorbati@yahoo.com}}}
\affiliation{Department of Physics, Tarbiat Modares University, Tehran, Iran}

\author{S. A. Jafari{\footnote {Electronic address: jafari@sharif.ir}}}
\affiliation{Department of Physics, Isfahan University of Technology, Isfahan 84156-83111, Iran}
\affiliation{Department of Physics, Sharif University of Technology, Tehran 11155-9161, Iran}
\affiliation{School of Physics, Institute for Research in Fundamental Sciences (IPM), Tehran 19395-5531, Iran}

\begin{abstract}
We use Perturbative Continuous Unitary Transformations (PCUT) to study 
the one dimensional Extended Ionic Hubbard Model (EIHM) at half-filling in the band insulator 
region. The extended ionic Hubbard model, in addition to the usual ionic
Hubbard model, includes an inter-site nearest-neighbor (n.n.) repulsion, $V$. We consider 
the ionic potential as unperturbed part of the Hamiltonian, while the hopping and interaction 
(quartic) terms are treated as perturbation. 
We calculate total energy and ionicity in the ground state. 
Above the ground state, (i) we calculate the single particle excitation spectrum by adding
an electron or a hole to the system.  (ii) the coherence-length 
and spectrum of electron-hole excitation are obtained. 
Our calculations reveal that for $V=0$, there are two triplet bound state modes and three 
singlet modes, two anti-bound states and one bound state, while for finite values 
of $V$ there are four excitonic bound states corresponding to two singlet and two triplet modes.
The major role of on-site Coulomb repulsion $U$ is to split singlet and triplet
collective excitation branches, while $V$ tends to pull the singlet branches below
the continuum to make them bound states.
\end{abstract}
\maketitle

\section{Introduction}

The ionic Hubbard model was introduced to describe the 
Neutral-Ionic transition in the charge-transfer organic 
mixed-stack compounds such as TTF-Chloranil~\cite{Hubbard}. The
transition occurs by change in pressure or temperature 
of these materials~\cite{Torrance1,Torrance2}. Since then, 
in addition to experimental studies~\cite{Tokura,Mitani}, some theoretical investigations 
for identifying the nature of possible phases and optical properties 
have been conducted on this model~\cite{Nagaosa1} and its extended versions~\cite{Nagaosa2,Avignon}.

 On the other hand, in transition metal oxides BaTiO$_3$, KNbO$_3$, and KTaO$_3$  
it was noted that the valence of the transition metal ion is $\sim +3$~\cite{Cohen,Neumman}, 
rather than $+4$ which is expected if a simple ionic picture holds. Therefore it is necessary to consider
effects of strong correlations in these materials, leading to the so called 
ionic Hubbard model (IHM)~\cite{Egami}. 
Although the phase transitions at zero~\cite{Fabrizio,Batista,Manmana,Garg,Kampf,Hafez1} 
and finite~\cite{Paris,Bouadim,Hafez2} temperatures for this model 
have been studied by many researchers, the physical properties of 
band-insulating regime of this model has not been studied much~\cite{Maeshima}.

Besides Hubbard-Peierls Hamiltonian~\cite{Ono}, the ionic Hubbard model is also 
a candidate for Halogen-bridged transition-metal compounds~\cite{Maeshima} 
known as MX-chains, where M stands for transition-metal and X for Halogen. 
There are many theoretical and experimental researches on the optical properties 
of these compounds. In NiX-chains, due to large on site interaction, 
the system is Mott insulator while PdX- and PtX-chains are band insulators~\cite{Batistic,Iwai,Ohara}.

  The one dimensional EIHM Hamiltonian is as follows:
\bearr
H &=& \frac{\Delta}{2}\sum_{i,\sigma}(-1)^{i}n_{i,\sigma}+
\frac{t}{4}\sum_{i,\sigma}(c^{\dagger}_{i,\sigma}c_{i+1,\sigma}+h.c.)\nn\\&&+
U\sum_{i}n_{i,\uparrow}n_{i,\downarrow}+\frac{V}{2}\sum_{i}
n_{i}n_{i+1},
\label{EIHM}
\eearr
where $\Delta$ is one-particle ionic potential, $t/4$ is nearest-neighbor hopping amplitude, 
$U$ represents on site repulsive interaction, and $V/2$ is nearest-neighbor inter-site interaction. 
$c_{i,\sigma}^{\dagger}$ and $c_{i,\sigma}$ are electron creation and annihilation 
operators at site $i$ with spin $\sigma$, respectively. $n_{i,\sigma} \equiv 
c_{i,\sigma}^{\dagger}c_{i,\sigma}$ is the electron occupation operator and 
$n_i \equiv \sum_{\sigma}n_{i,\sigma}$. 
When $\Delta$ is dominant energy scale, i.e. $\Delta \gg t, U, V$ the ground state 
of this model is a band insulator, with electrons and holes and their possible (anti)bound-states 
constituting the excitation spectrum. When $U$ dominates, i.e. $U \gg t,\Delta, V$ the
ground state is Mott insulator, with doublon and holon excitation. 
In this paper, we investigate the one dimensional EIHM by PCUT method
in the band insulator region. We take the first term in Eq.~(\ref{EIHM}) 
as unperturbed Hamiltonian, and the rest as perturbation. 
In the following, PCUT method is introduced. Then by using this 
method, ground state energy, ionicity, spin and charge gaps, one electron and one hole dispersion, 
as well as spectrum and coherence length of the electron-hole excitation for EIHM are calculated.

\section{PCUT method}
  PCUT or perturbative flow equation method is a high order perturbative 
approach for calculating the multi-particle excitations and observables
in model Hamiltonians~\cite{Knetter1,Knetter2,Knetter3,Fischer}. 
In this method, first the Hamiltonian is mapped, by flow equations 
approach~\cite{Wegner}, to a perturbative effective Hamiltonian. 
This effective Hamiltonian is obtained using a generator\cite{Mielke} 
which conserves the number of quasi-particles that are the excitations of 
unperturbed part. Then by using  
linked cluster expansion theorem~\cite{Gelfand}, physical quantities 
in the thermodynamic limit can be calculated employing  finite size clusters. 
To be self-contained, here we concisely review PCUT method.

  Assume that the Hamiltonian can be written as a perturbative problem:
\be
H=Q+xT,
\label{BPH}
\ee
where $xT$ is the part of the Hamiltonian, the effect of which we would like to 
consider perturbatively. In order to apply PCUT method on the Hamiltonian~(\ref{BPH}), 
$Q$ is required to have equi-distance spectrum. The difference between two successive 
levels is chosen as the unit of energy. Also $T$ must be of the following form:
\be
T=\sum_{n=-N}^{+N}T_{n},
\label{T}
\ee
where $N$ is a finite number, and $T_{n}$ is a ladder operator which decreases or 
increases the eigen-values of $Q$ by $n$ units.

When these conditions are satisfied, the flow equation and the quasi-particle conserving
generator respectively read,
\be
\partial_\ell H(\ell)=[\eta(\ell),H(\ell)],
\ee
\be
\eta_{i,j}(\ell)={\rm{sign}}(Q_i-Q_j)H_{i,j}(\ell),
\ee
where $Q_i$ is the $i$th eigen value of $Q$. These equations can be used to transform 
the Hamiltonian~(\ref{BPH}) to the following effective Hamiltonian,
\be
H_{\rm eff}=Q+\sum_{k=1}^{\infty}x^{k}\sum_{\mid{\overrightarrow{m}\mid=k},M(\overrightarrow{m})=0}
C(\overrightarrow{m})T(\overrightarrow{m}),
\label{EH}
\ee
where $\vert \overrightarrow{m} \vert =k$ means that $\overrightarrow{m}$ is a vector of order $k$, 
$T(\overrightarrow{m})=T_{m_{1}} \cdots T_{m_{k}}$, $M(\overrightarrow{m})=m_{1}+\cdots+m_{k}$, and 
the summation is over the components of $\overrightarrow{m}$ which can take values $-N,-N+1\cdots +N$. 
The $C(\overrightarrow{m})$ are known as fractional coefficients which can be obtained from a set of coupled nonlinear 
first order differential equations. Uhrig and his group, who formulated PCUT method, 
calculated these coefficients up to high orders which are available, electronically~\cite{Uhrig}. 
This part of the method is quite generic and holds for all Hamiltonians 
that fulfill the above conditions. The salient feature of the PCUT effective 
Hamiltonian is the constraint $M(\overrightarrow{m})=0$. This constraint causes 
the effective Hamiltonian to connect only the states of $Q$ corresponding to same eigen-values. 
Therefore, interpreting the excitations of $Q$ as quasi-particles (QPs), it is sufficient 
to diagonalize the effective Hamiltonian only in sectors with definite number of
quasi-particles, e.g. in 0-, 1-, and 2-particle sectors. This provides a picture
of the low lying excitation spectrum of the system.

  In order to calculate the matrix elements of $H_{\rm eff}$ for an 
infinite lattice, first it is necessary to decompose 
the effective Hamiltonian to irreducible $n$-particle operators that are 
cluster additive~\cite{Knetter2}. Only for these operators, the matrix
elements calculated for finite size are valid in thermodynamic limit, i.e.
there would be no finite size error. Therefore in the next step one needs to identify minimum 
cluster size along with suitable form of boundary conditions to calculate required matrix elements~\cite{Knetter3}.
Note that despite the generic part of the method, the choice of cluster depends on 
the properties of the specific Hamiltonian at hand. 

PCUT method has been extensively used in spin systems in order to 
calculate ground state energy, 1-particle dispersion, 2-particle excitation 
spectrum, and observables~\cite{Arlego,Brenig,Schmidt1,Knetter4,Schmidt2,Vidal,Dusuel}. 
However, to our knowledge, this method has not been 
applied much to fermionic systems~\cite{Yang}. In this paper 
the aim is to implement this method on EIHM Hamiltonian in order 
to calculate ground state energy, 1-electron and 1-hole dispersions, as well as spectrum and 
coherence-length of electron-hole excitation. We use Pad\'{e} approximation with polynomials of various orders to enlarge the convergence radius of the obtained perturbative expansions. 

\section{Application of PCUT method to EIHM}
  To implement PCUT method on EIHM, we write this Hamiltonian as Eq.~(\ref{BPH}). 
Now we identify $x=t$ and,
\bearr
Q&=&\frac{\Delta}{2}\sum_{i,\sigma}(-1)^i n_{i,\sigma},
\label{Q}
\\
T &=& \frac{1}{4}\sum_{i,\sigma}(c^{\dagger}_{i,\sigma}c_{i+1,\sigma}+h.c.)+
u\sum_{i}n_{i,\uparrow}n_{i,\downarrow} \nn \\&&+\frac{v}{2}\sum_{i}
n_{i}n_{i+1},
\label{PP}
\eearr
where $u\equiv\dfrac{U}{t}$, and $v \equiv\dfrac{V}{t}$. {\em It is convenient to choose $\Delta$ as energy 
unit and set $\Delta=1$}. 
Note that the popular convention for the unit of energy in Hubbard model literature
is to take $t$ as unit of energy. Moreover, a different sign convention for $t$ is used
in the literature. This should be noted when comparing our results
with other references. 
It is obvious that at half-filling, $Q$ has an equi-distance spectrum and its energy 
difference between two successive levels is $1$. Therefore the first condition is satisfied. 
The exact ground state of $Q$ corresponds to a configuration where 
all odd sites are doubly occupied and all even sites are left empty.
Since there are no electrons in even sites, nor holes in odd sites,
in the sense of having no quasi-particles, one can interpret the ground state as {\em vacuum}. 
The excitations of $Q$ can be obtained by moving an electron from odd site to an even site.  Or
alternatively in terms of quasi-particle picture, when an electron is added to even 
sites and simultaneously a hole is added to odd sites. This 
is a 2-particle excitation that includes two types of particles. For obtaining the 
properties of $H_{\rm eff}$ in this sector we must first discuss 1-particle sectors 
corresponding to a single electron (hole) added to even (odd) sites. 
These one particle sectors themselves include very important physical information like 
electron and hole dispersions, which can be directly measured in (inverse) photo-emission
spectroscopies. 

  Since the eigen-states of $Q$ can be used as basis for $H$, it is possible 
to write $T$ in the form of Eq.~(\ref{T}). The interaction terms in Eq.~(\ref{PP}) compose 
$T_{0}$ and with some simple calculations we can decompose hopping term to $T_{-1}$, 
and $T_{+1}$. Therefore second condition is satisfied with $N=1$, which 
reflects the fact that hopping is associated with a pair of particle-holes. 
These operators are defined as $T_n=\sum_{i}\Gamma_n(i)$, where summation is over lattice points 
and $\Gamma_n(i)$ only connects nearest-neighbor sites, i.e. $i$ and $i+1$. 
The $\Gamma_n(i)$ are defined as:
\be 
\Gamma_0(i)=un_{i,\uparrow}n_{i,\downarrow}+\frac{v}{2}n_in_{i+1},
\label{Gamma0}
\ee
\bearr
4\Gamma_{+1}(i)&=&|0,\uparrow \rangle \langle \uparrow, 0| 
-|\uparrow ,d\rangle \langle d,\uparrow|
-|\downarrow ,d\rangle \langle d, \downarrow| \nn \\
&+& |0,d \rangle \langle \uparrow, \downarrow| 
-|\downarrow ,\uparrow \rangle \langle d,0| 
+|\uparrow ,\downarrow \rangle \langle d,0| \nn \\
&+&|0,\downarrow \rangle \langle \downarrow ,0|
-|0,d\rangle \langle \downarrow, \uparrow| ~~~~~,~~~~ i \in {\rm odd},
\label{Gamma+1}
\eearr
where $|a,b\rangle$ in the above notation means $|a\rangle_i \otimes |b\rangle_{i+1}$, and 
$|0\rangle$, $|\downarrow \rangle$, $|\uparrow \rangle$, and $|d \rangle$ stand for a site occupied with no, spin down, spin up, and two electron(s), respectively. The above term is essentially a right to left hopping,
which has been expanded in terms of Hubbard operators for practical purposes.
We also have $\Gamma_{+1}(i \in {\rm even})=\Gamma_{+1}^{\dagger}(i \in {\rm odd})$, 
$\Gamma_{-1}(i \in {\rm even})=\Gamma_{+1}(i \in {\rm odd})$, and $\Gamma_{-1}(i \in {\rm odd})=\Gamma_{+1}(i \in {\rm even})$.
In writing Eq.~(\ref{Gamma+1}) we have used the following sign convention for 
Fermi operators: When $c_{i,\sigma}$ or $c_{i,\sigma}^{\dagger}$ act on a state 
a phase factor $(-1)^{S_{i,\sigma}}$ is produced, where $S_{i,\sigma}$ denotes
the total (both $\up$ and $\down$) number of electrons located in the left side of site $i$. 
Moreover, at any given site $i$, spin-$\up$ electron always appears to the left side of 
spin-$\down$ electron. As long as one is interested in one dimensional lattices the choice 
of suitable clusters to find the irreducible interactions is simple while it becomes 
a tricky task in higher dimensions and for lattices with complicated structures. 
It is obvious from Eqs.~(\ref{Gamma0}) and~(\ref{Gamma+1}) that  
when $\Gamma_{\pm1}(i)$ acts on two nearest-neighbor sites, it always transfers 
the quasi-particles. But $\Gamma_0(i)$ acts on two nearest-neighbor sites 
with no transfer of QPs. When we calculate 1-particle hopping terms and 2-particle 
interactions in thermodynamic limit based on full graph theory\cite{Knetter3}, 
this property of $\Gamma_0(i)$ in EIHM, in analogy with models involving $4$-spin 
interactions\cite{Schmidt3}, generates larger clusters compared to systems with 
solely $2$-spin interactions perturbed around the dimer limit~\cite{Knetter1,Knetter3}.

\subsection{Ground state energy and ionicity}
\begin{figure}[t]
  \begin{center}
    \includegraphics[angle=-90,width=8cm]{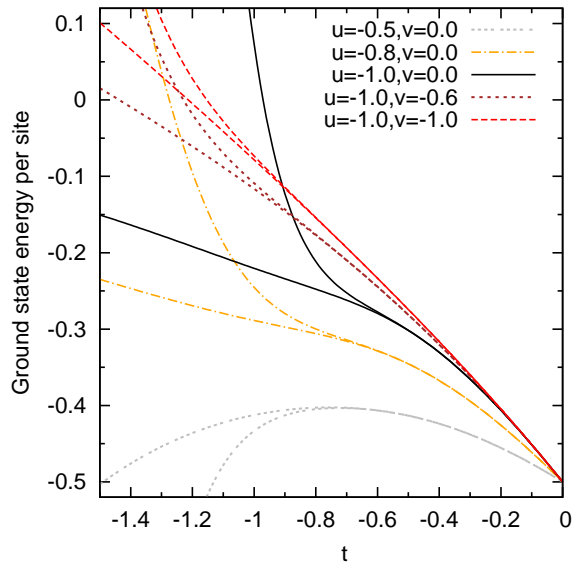}
    \caption{Ground state energy per site versus $t$ 
    for $\Delta=1$ and different values of $u$ and $v$. 
    In each parameter set both plain series and the best 
    Pad\'{e} approximation have been depicted. The plain series 
    expansions diverge at some special values of $t$ while the Pad\'{e}
    approximant are converged in the plotted region.}
    \label{E0}
  \end{center}
\end{figure}
\begin{figure}[t]
  \begin{center}
    \includegraphics[angle=-90,width=8cm]{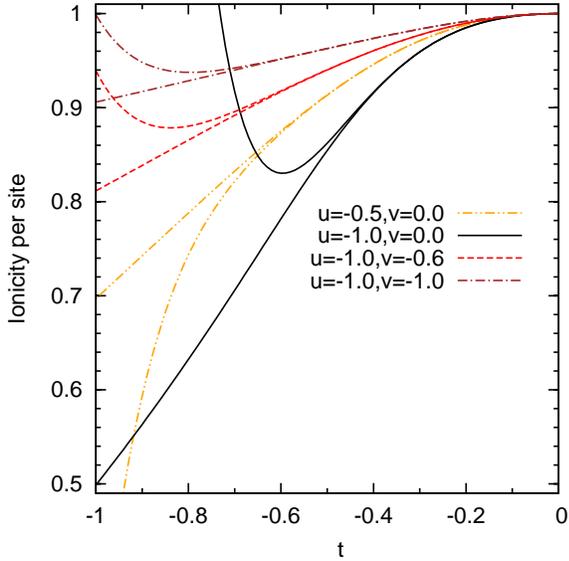}
    \caption{Ionicity per site versus $t$ for $\Delta=1$ 
    and different values of $u$ and $v$. In each case two graphs 
    corresponding to the best 
    Pad\'{e} approximation and the plain series have been plotted. The 
    plain series expansions diverge at some special values of $t$. 
    The ionicity per site decreases (increases) by increasing in $U$ ($V$).
    }
    \label{ionicity}
  \end{center}
\end{figure}
Since $H_{\rm eff}$ and $Q$ have the same ground state (unless a phase 
transition takes place)~\cite{Heidbrink} and since this 
state is not degenerate, the ground state energy of $H$ per site is given by 
$\epsilon_0=\frac{1}{L}\langle 0 |H_{\rm eff}|0 \rangle$, where $|0 \rangle$ is the ground state of $Q$ 
or the complex vacuum. We have calculated $\epsilon_0$ up to order $9$ in $t$, $U$, and $V$. 
At this order, from Eqs.~(\ref{Gamma0}) and (\ref{Gamma+1}) it turns out that
the minimum cluster size is $10$ with periodic boundary condition to obtain 
the thermodynamic limit value of $\epsilon_0$. Ground state energy per site  
is given in appendix. We have plotted ground state energy per site
versus $t$ for different values of dimensionless interaction parameters $u$ and $v$ in Fig.~\ref{E0}. 
Note that in this figure, each graph represents the best Pad\'e approximation. 
It is seen form Fig.~\ref{E0} that ground state energy 
increase by increasing $U$ and $V$. This is physically plausible, as 
repulsive interactions always increase energy of the system, unless a phase transition 
occurs. Note that since in our convention, electron-like bands are
represented with negative values of $t$, the negative values of $u,v$ in 
Fig.~\ref{E0} correspond to repulsive interactions $U,V>0$.

 From the analytical expression for ground state energy, we 
can use the Hellman-Feynman theorem to calculate the 
ionicity per site which is an important quantity for the ionic Hubbard 
model and is given by\cite{Manmana}:
\be
I \equiv -\frac{1}{L}\sum_{i,\sigma}(-1)^in_{i,\sigma}=-2\frac{\partial \epsilon_0}{\partial \Delta}.
\label{I}
\ee
The ionicity per site versus $t$ for different values of $u$ and $v$ 
has been plotted in Fig.~\ref{ionicity}. In this figure, only the best Pad\'{e} 
approximation have been depicted. 
This figure shows that as $U$ is increased 
the ionicity decreases while increment in $V$ increases the ionicity. 
Therefore the on-site Coulomb interaction $U$ tends to decrease the valence.
However, the next n.n. term $V$ has the opposite effect. 
This behavior of $V$ is expected, as it stabilizes configurations in which
n.n. sites have opposite charges with respect to the half-filled background; i.e.
one is doubly occupied, with its n.n. site being empty.

  If we regard Eq.~(\ref{epsilon0.eqn}) for $\epsilon_0$ as a 
series expansion in $t$, the coefficient of $t^{2}$ 
can be summed as a geometric series, such that the ground state energy 
per site up to order $t^{2}$ in band insulator phase is,
\be
\epsilon_0=\dfrac{U-\Delta}{2}-\dfrac{1}{8}\dfrac{t^{2}}{\Delta+
\dfrac{3}{2}V-U}+O(t^{4}).
\label{GSGt2}
\ee
Convergence radius for this expression is obviously given by $\Delta>|3V/2-U|$. 
This expression can be interpreted as follows: In the band insulating
regime, $\Delta\gg U,V$, when $t=0$ the ground state belongs to 
subspace of configurations in which odd sites are doubly occupied. 
By turning on the perturbation $t$, a second order energy gain 
can be obtained when an electron virtually hops to even neighboring site.
From above expression and Eq.~(\ref{I}), we can obtain the ionicity per site 
at the same order as follows,
\be
I=1-\dfrac{1}{4}\dfrac{t^2}{(\Delta+\dfrac{3}{2}V-U)^2}+O(t^4).
\ee
These are perturbative expressions for ground state energy and ionicity  
around atomic limit that are valid for $t\ll \Delta+\frac{3}{2}V-U$
in band insulating phase. In the Mott insulating region, similar expressions 
can be obtained for these quantities at $V=0$ in the limit $t\ll U-\Delta$~\cite{Manmana}. 
These expressions are derived based on the mapping of the ionic Hubbard model to 
the Heisenberg model~\cite{Nagaosa1} whose exact ground state is known in 1D.

\subsection{Electron and hole dispersions}


  To calculate one-particle dispersion, one must add 
an electron to the even sites of the system, while the
corresponding hole must be added to odd sites.
Since $H_{\rm eff}$ only connects those states of $Q$ 
which have same eigen-values, it can merely transfer 
the electron between even sites and the hole between odd sites. 
Therefore it generates hopping terms of the form 
{\setlength\arraycolsep{1pt}
\bearr
   H_{\rm eff} &= 
   & E_0+t_{0,e} \sum_{i\sigma} e^\dagger_{i\sigma}e_{i\sigma}+
   \sum_{r=2,4,\ldots} t_{r} \sum_{i\sigma}  e^\dagger_{i,\sigma}e_{i+r,\sigma} + \mbox{h.c.} \nn \\
   &+& t_{0,h} \sum_{i\sigma} h^\dagger_{i\sigma}h_{i\sigma}+
   \sum_{r=2,4,\ldots} t_{r} \sum_{i\sigma}  h^\dagger_{i,\sigma}h_{i+r,\sigma} + \mbox{h.c.},
\eearr}
where $E_0$ is the ground state energy and $e^\dagger_{i\sigma}$ ($h^\dagger_{i\sigma}$) creates an electron (hole)
with spin $\sigma$ at even (odd) site $i$.
The long range hopping amplitudes $t_r,r=2,4,\ldots$ for electron and hole
are identical up to all orders in perturbation theory. 
However, as for the shortest range hopping amplitude, $t_0$ (i.e. no hopping) 
only the first order of perturbative expansion for the electrons and holes are 
different and are related by $t_{0,e}-\mu=t_{0,h}+\mu$,
where $\mu=U/2+V$ is the chemical potential at half-filling for Eq.~(\ref{EIHM}). 
Note that if we had started from a particle-hole symmetric version of the Hamiltonian,
by replacing $n_{i\sigma}\to n_{i\sigma}-1/2$ in $U$ and $V$ terms of
Eq.~(\ref{EIHM}), we would have $\mu=0$ leading to $t_{0,e}=t_{0,h}$. 

The hopping coefficients $t_r$
in the thermodynamic limit can be calculated on a finite 
cluster based on linked cluster theorem~\cite{Gelfand}. 
To calculate these coefficients, first we must determine the minimum 
cluster size which can give us the thermodynamic limit results. 
The minimum cluster size, in addition to order of perturbation, depends on 
the particular hopping coefficient which we would like to calculate. 
For example, to obtain $t_0$ up to order $9$, $13$ sites and up to order $10$, $15$ sites are needed. 
For calculation of $t_2$ up to order $9$, $10$, and $11$ 
one needs $13$, $13$, and $15$ sites, respectively. 
For $t_4$, $t_6$, $t_8$, up to order $9$, $11$ sites are 
sufficient to calculate them in thermodynamic limit. 
These numbers can be inferred from Eqs.~(\ref{Gamma0}) and (\ref{Gamma+1}). 
We have calculated electron dispersion up to order $9$. 
At this order there are five hopping coefficient, $t_0$, $t_2$, $t_4$, 
$t_6$, and $t_8$ which are independent of the spin of the electron and 
have been reported in appendix. 

Electron and hole dispersions of the effective Hamiltonian are given by,
\be
\omega_{e(h)}(k)=\pm(t_{0,e(h)}+2\sum_{n=1}^{4}t_{2n}\cos(2nk)).
\ee
\begin{figure}[t]
  \begin{center}
    \includegraphics[angle=-90,width=8cm]{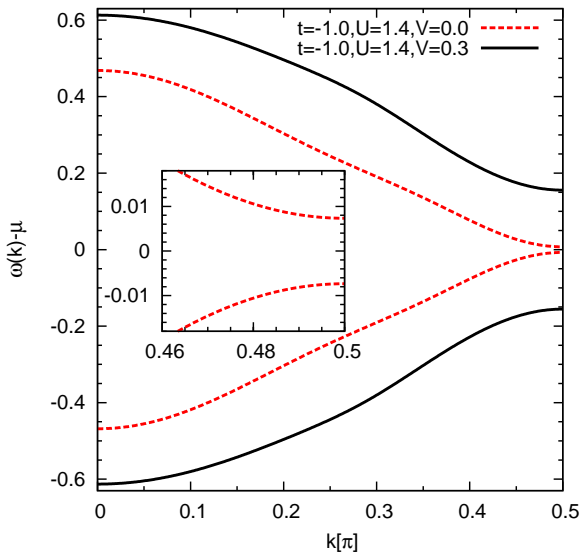}
    \caption{Electron and hole dispersions around Fermi surface $\mu=\frac{U}{2}+V$ 
    for ($t=-1.0$, $U=1.4$, $V=0$) and ($t=-1.0$, $U=1.4$, $V=0.3$) at $\Delta=1$. We have used 
    Pad\'{e}[7,2] in both cases. Our result for charge gap at ($t=-1.0$, $U=1.4$, $V=0$) 
    is $0.014$ which is in excellent agreement with DMRG result that is $0.012$. 
    }
    \label{disp2}
  \end{center}
\end{figure}

\begin{figure}[t]
  \begin{center}
    \includegraphics[angle=-90,width=8cm]{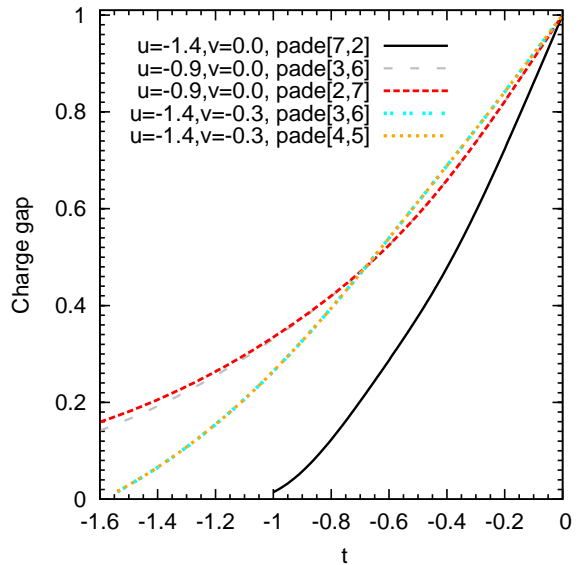}
    \caption{Charge gap versus $t$ for different values of 
    $u$ and $v$ in band insulator region. In all cases Pad\'{e} approximation $[p,q]$
    corresponding to quotient of polynomials of order $p,q$ has been used, which
    is indicated in the legend. Spin and charge gaps are equal, 
    because the hopping coefficients, i.e. $t_n$, are independent of spin.
    }
    \label{CGF}
  \end{center}
\end{figure}
In Fig.~\ref{disp2} we have plotted the one-particle dispersion corresponding to 
$t=-1.0$, $U=1.4$, $V=0$, as well as $t=-1.0$, $U=1.4$, $V=0.3$. 
Note that to obtain reliable results for the value of $U=1.4$ we have used Pad\'{e} 
approximation. The charge gap is defined as,
\be
E_{c} \equiv E_{0}(N+1)+E_0(N-1)-2E_0(N).
\ee
For $U=0,V=0$, we have a simple band insulator with gap magnitude $\Delta$ which
has been taken to be unit of energy. In Fig.~\ref{disp2} the charge gap in terms 
of the effective Hamiltonian becomes, $\omega_e(k=\frac{\pi}{2})-\omega_h(k=\frac{\pi}{2})$.
As can be seen for $t=-1,U=1.4, V=0.0$, the gap becomes very small $\approx 0.014$, 
which agrees very well with corresponding DMRG value~\cite{Manmana} $0.012$.
Note that when comparing results of Ref.~\onlinecite{Manmana} with ours, one has
to appropriately re-scale the Hamiltonian parameters. Beyond $U\approx 1.4$ the
band insulating phase terminates and one may have different behaviors for
spin and charge gaps~\cite{Manmana}. Although our perturbative treatment does not
extend beyond the band insulating state, extension of the convergence radius
captures a critical value of $U$ at which the charge gap tends to zero.
Now at $U=1.4$, we turn on $V$. As can be seen in Fig.~\ref{disp2}, turning
on $V$ increases the gap.

Charge gap versus $t$ for different values of $u$ and $v$ have been 
plotted in Fig.~\ref{CGF}. For each value $u,v$ the best Pad\'{e} fits 
have been shown. For $u=-1.4$, $v=0$ the Pad\'e fit shown in the figure
approaches the DMRG value~\cite{Manmana} as $t\to -1$.
For other values of $u,v$ reported in this figure, to ensure the 
quality of results, for each value $(u,v)$ we have shown two 
Pad\'e fits. Comparing the variation of charge gap at $v=0$, as can be
seen by moving from $u=-0.9$ (red dashed line) to $u=-1.4$ (black solid line),
the gap magnitude decreases, which means Hubbard $U$ work against 
the band gap parameter $\Delta$ and tends to reduce it.
Similarly, for a fixed value of $u$, turning on the n.n. repulsion $V=vt$
increases the band gap.

Similar to ground state energy we find expressions for $t_{0,e}$, and 
$t_{0,h}$, around atomic limit up to order $t^2$ for $V=0$. 
These expressions are as follows:
\be
t_{0,e}=t_{0,h}+U=\frac{\Delta}{2}+\frac{t^2}{8} \frac{\Delta+U}{\Delta(\Delta-U)}+O(t^4),
\label{t0t2}
\ee
and for $t_2$ at finite $V$ up to order $t^2$ we obtain:
\be
t_2=\frac{1}{16} \frac{t^2}{\Delta+V-U}+O(t^4).
\label{t2t2}
\ee
From these expressions we can find the electron and hole dispersions 
for $V=0$. The dispersions are as $\omega(k)=t_0+2t_2\cos(2k)$. 
This gives rise to inverse square root density of states-- a characteristic
of 1D bands, the width of which is controlled by $t_2$.

In this subsection we obtained some informations about EIHM at half-filling by 
adding one electron and one hole to the system, separately. This gave rise
to the band dispersion relation of the system. Now let us study the particle-hole
(charge neutral) excitations of the system.

\subsection{Spectrum and coherence length of electron-hole excitation}
The first excited state of $Q$ at half-filling is when one electron from an odd site with 
energy $-\frac{1}{2}$ is transfered to an even site with energy $+\frac{1}{2}$ or 
in quasi-particle language when one electron is added to even sites and one hole 
is added to odd sites, simultaneously. This is a 2-particle excitation which include 
two types of particle, i.e. electron and hole. It is possible to 
discuss the spectrum and coherence length of this electron-hole excitation by PCUT method. 
State of the system in this sector can be represented as $|i,m_e;j,m_h\rangle$ 
where $i$ is the even site and $j$ is the odd site at which electron and hole 
excitations are created, respectively. Also $m_e$ and $m_h$ are magnetic quantum numbers 
of electron and hole. To discuss the two-particle sector, in addition to quadratic (hopping)
terms, one needs to include quartic electron-hole terms of the following form 
in the effective Hamiltonian,
$$
  t_{i_1,i_2;j_1,j_2} h^\dagger_{i_1} e^\dagger_{i_2} e_{j_2} h_{j_1}
$$
Since total $H_{\rm eff}$ preserves the total spin of the electron-hole pair, we must use the basis 
$|i,j;S,M\rangle$, where $S$ is total spin of electron-hole pair and $M$ is its magnetic quantum number. 
$S$ can be either $0$ or $1$ corresponding to singlet and triplet total spin states.
When $H_{\rm eff}$ acts on states with a particle-hole pair, it gives rise to a 
new configuration in the subspace with a particle-hole pair. The processes of mapping 
the initial state to final state can be classified as $2$-particle, 
a mixture of $1$- and $2$-particle, or a mixture of $0$-, $1$-, and $2$-particle process. 
In order to discuss this $2$-particle sector, first one has to determine 
the \textit{irreducible} 2-particle interactions obtained by subtraction of
$0$- and $1$- particle contributions which by linked cluster theory can be employed to 
calculate $t_{i_1,i_2;j_1,j_2}$ for a thermodynamics system on a finite cluster~\cite{Knetter2,Knetter3}. 
Again similar to 1-particle sector, the minimum cluster size 
can be identified from properties of Eqs.~(\ref{Gamma0}) and (\ref{Gamma+1}). 
We have calculated these coefficients up to order $8$ in $t,U,V$. 
These coefficients turn out to be solely dependent on total spin $S$ 
and not on magnetic quantum number $M$.
\begin{figure}[t]
  \begin{center}
    \includegraphics[angle=-90,width=8cm]{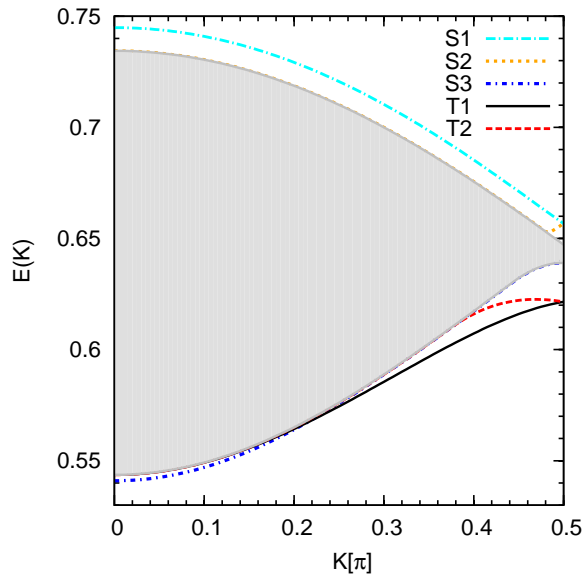}
    \caption{Plain series result for electron-hole energy versus $K$ for $t=-0.5$, $U=0.5$, and $V=0$. 
    There are three singlet bound/anti-bound state branches and two triplet bound state 
    branches. The two singlet anti-bound state modes and two triplet bound state 
    modes are degenerate at $K=\pm\frac{\pi}{2}$.
    }
    \label{2PEV=0}
  \end{center}
\end{figure}

When the interaction terms are identified, we have only 
a 2-particle problem that must be solved. Such a problem can be solved by changing the 
basis to total particle-hole momentum and their relative distance $|K,d\rangle$, or to total momentum and 
relative momentum $|K,K'\rangle$ of particle-hole pair~\cite{Zheng}. 
We use $|K,d\rangle$ representation, as in this basis it is 
possible to explicitly calculate electron-hole coherence length defined as~\cite{Zheng}:
\be
   L=\dfrac{\sum_d f_d^2 \vert d \vert}{\sum_d f_d^2},
   \label{CLE}
\ee
where $f_d$ is the probability amplitude for finding the particle-hole pair 
at a relative distance $d$. In our case $d$ can take only odd values as
electron are always in even sites and hole in odd sites.
\begin{figure}[t]
  \begin{center}
    \includegraphics[angle=-90,width=8cm]{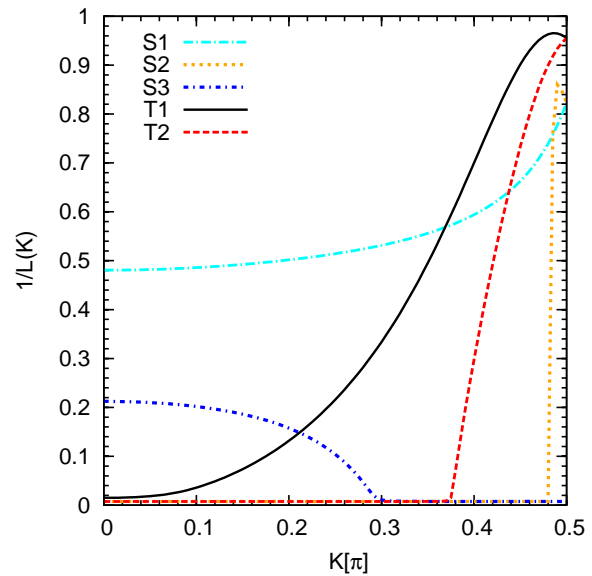}
    \caption{Inverse of coherence length versus $K$ for $t=-0.5$, $U=0.5$, and $V=0$. 
    For $S1$ mode the coherence length remains finite in the whole of range of momenta. 
    But for other modes it becomes infinite at some values of $K$. Coherence 
    length of $S2$ has a discontinuity at $K^*=0.46\pi$.
    }
    \label{CLFV=0}
  \end{center}
\end{figure}

Now that we have the explicit form of the $H_{\rm eff}$ in 2-particle sector, 
it remains to diagonalize $H_{d,d'}(K)$ for a given value of center of mass momentum $K$.
Here $d,d'$ indicate relative distance. 
For problems where the two particles are of same type (e.g. in Cooper pairing channel), 
this matrix is a semi-infinite~\cite{Knetter3}, but for problems with two types of particles, 
such as our case (i.e. in particle-hole channel), it is an infinite matrix. 
However, when dealing with these matrices on the computer, one constructs 
finite $N\times N$  dimensional representations of the Hamiltonian for which 
the eigen-value problem must be solved. Next by systematically increasing $N$,
the spectrum at $N\to\infty$ limit can be obtained.
For those values of $K$ where the particle-hole pairs form a bound or anti-bound state,
the convergence is achieved quickly. This numeric observation is physically plausible,
because for split-off states, the two particles are close to each other and hence the
finite size effects become irrelevant. By determining eigen-vectors of the matrix we 
obtain the $f_d$ which can be used to calculate the coherence length via Eq.~(\ref{CLE}).

Upper and lower edges of the particle-hole continuum (PHC) is obtained by ignoring 
the 2-particle interactions and diagonalizing $H_{\rm eff}$ in the basis of $|K,K'\rangle$
which gives rise to {\em free} particle-hole energies. Then 
the maximum (minimum) eigen-value in $K'$ for a fixed value of $K$ in Brillouin zone 
determines the upper (lower) edge of the PHC~\cite{Knetter3,Zheng}.

Plain series result of electron-hole excitation spectrum for $t=-0.5$, $U=0.5$, and 
$V=0$ is depicted in Fig.~\ref{2PEV=0}. This figure 
shows that there are three singlet and two triplet modes. The singlet channel contains  
two anti-bound states and one bound state mode while the triplet modes are bound states. 
The two singlet anti-bound state branches and the two triplet branches are degenerate 
at the edges of the Brillouin zone, $K=\pm\frac{\pi}{2}$. Hence, when 
longer range Coulomb interaction ($V$) is absent, on-site interaction $U$ 
stabilizes the triplet excitations with respect to singlet ones. Note 
that in the limit $U\to 0$, a simple band picture would imply the
singlet and triplet branches to have the same excitation energies.
Therefore such singlet-triplet splitting is a remarkable effect of
short range correlations which manifests itself at molecular~\cite{daSilva}
as well as infinite size systems~\cite{BaskaranJafari} where the correlation
effects is believed to play important role.

\begin{figure}[t]
  \begin{center}
    \includegraphics[angle=-90,width=8cm]{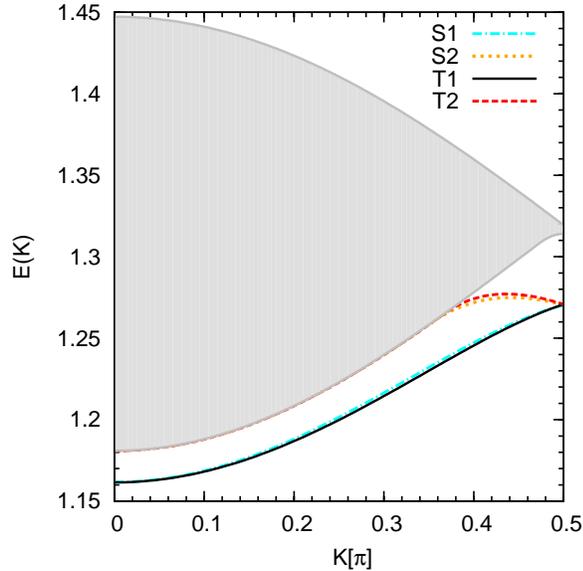}
    \caption{Plain series result of electron-hole energy versus $K$ for $t=-0.8$, $U=0.0$, and $V=0.1$. 
    There are two singlet ($S1$ and $S2$) and two triplet ($T1$ and $T2$) bound state 
    modes. Each singlet is nearly degenerate with one triplet which show the effect 
    of on site interaction $U$ in separation of singlet and triplet state.
    }
    \label{2PEU=0}
  \end{center}
\end{figure}

We have also plotted the inverse of the
coherence length in Fig.~\ref{CLFV=0} corresponding to parameter values  
of Fig.~\ref{2PEV=0}. Except for $S1$ and $S2$ branches, the coherence length increases with 
decreasing  bounding energy. Coherence lengths of $S1$ and $S2$ decrease 
with decreasing of anti-bounding energy and $S2$ mode has a discontinuity at $K^*=0.46\pi$. 
For $S1$ mode which is anti-bound everywhere in the BZ, the coherence length 
always remains finite. But for other branches who enter the PHC at some momentum, 
the coherence length become infinite at those particular momenta.

\begin{figure}[bth]
  \begin{center}
    \includegraphics[angle=-90,width=8cm]{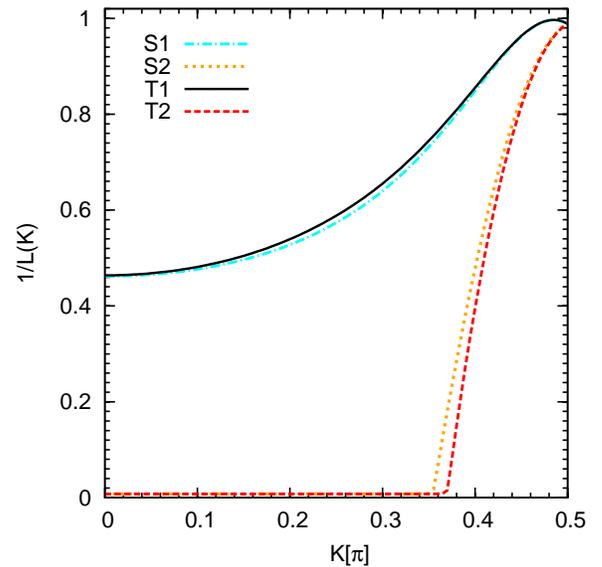}
    \caption{Inverse of the coherence length versus $K$ for $t=-0.8$, $U=0.0$, and $V=0.1$. 
    Inverse of coherence length is identical for $S2$ and $T2$ and increases monotonically 
    with $K$ while for $S1$ and $T1$ there is a peak not exact at the border.
    }
    \label{CLFU=0}
  \end{center}
\end{figure}

Plain series results for electron-hole energy and inverse of coherence length versus $K$ for $t=-0.8$, $U=0.0$, and $V=0.1$ 
have been depicted in Figs.~\ref{2PEU=0} and \ref{CLFU=0}, respectively. Fig.~\ref{2PEU=0} 
shows that there are two singlet and two triplet modes which all of them are bound states. 
The fact that the all modes are bound state compared with previous case, $V=0$, 
indicates that the long range part of the Coulomb interaction amounts to {\em attraction} 
among electron-hole pairs in both singlet and triplet channels which in turn forms the so
called excitonic states. 
Each of singlet branches is nearly degenerate with the corresponding triplet branch.
This again emphasizes that the short range Hubbard interaction $U$ is the major mechanism
behind the splitting between singlet and triplet collective mode branches. 

In Fig.~\ref{CLFU=0} we display the behavior for coherence length for the same parameter values
as in Fig.~\ref{2PEU=0}. Coherence length of $S2$ and $T2$, that are nearly degenerate, 
monotonically increases by decreasing $K$ until a certain value $K^*$ is reached, beyond
which the coherence length becomes infinite. 
But for $S1$ and $T1$ there is always a finite coherence length, indicating the 
corresponding p-h bound state exists everywhere in the BZ. 

\begin{figure}[t]
  \begin{center}
    \includegraphics[angle=-90,width=8cm]{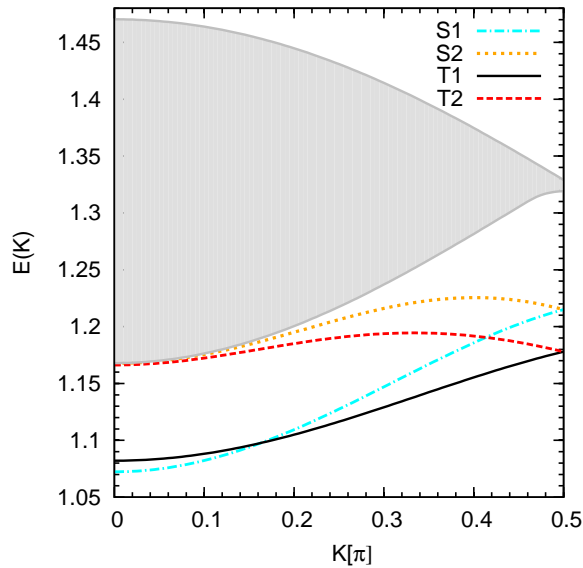}
    \caption{Plain series result of electron-hole energy versus $K$ for $t=-0.8$, $U=0.4$, and $V=0.3$. 
    There are two singlet, $S1$ and $S2$, and two triplet, $T1$ and $T2$, bound state 
    modes. For $K<0.16$ the lowest energy mode is singlet while for $K>0.16$ it is a 
    triplet state. At the border of Brillouin zone, $K=\pm \frac{\pi}{2}$, the two singlets and two triplets 
    are degenerate.
    }
    \label{2PE}
  \end{center}
\end{figure}

\begin{figure}[t]
  \begin{center}
    \includegraphics[angle=-90,width=8cm]{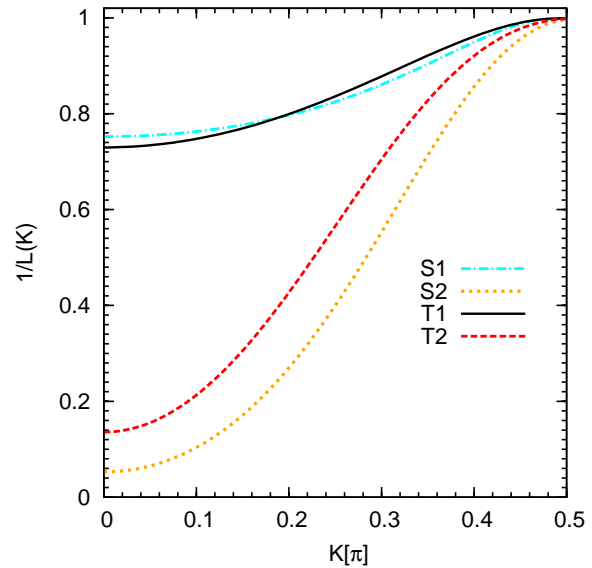}
    \caption{Inverse of coherence length versus $K$ for $t=-0.8$, $U=0.4$, and $V=0.3$. 
    There are four graphs correspond to two singlets, $S1$ and $S2$, and two triplets, 
    $T1$ and $T2$, modes. At the $K=\pm \frac{\pi}{2}$ all states have same coherence 
    length that is equal to $1$.
    }
    \label{CLF}
  \end{center}
\end{figure}

Now let us consider a case where both $U$ and $V$ are present.
In Figs.~\ref{2PE} and \ref{CLF}, we have plotted the plain series results for particle-hole energy 
and inverse coherence length versus $K$ for $t=-0.8$, $U=0.4$, and $V=0.3$. 
Again the behavior of coherence length and binding energy are in accordance with each other.
As can be see while due to presence of non-zero $V$, all modes are binding, still on-site
correlation $U$ leads to singlet-triplet splitting as expected. 
As can be seen in  Fig.~\ref{2PE}, near the BZ edge, triplets are lowest lying excitations.
But for smaller momentum transfer $K$, there are regions where the singlet mode
will become lowest excited state. 

In the appendix we give analytic expressions for the electron-hole energy 
$E_{\rm ph}(K)$ for singlet and triplet modes at the zone edge, $K=\frac{\pi}{2}$, 
up to order $8$. These relations are valid for $t^8 \ll V$.

\section{summary discussion} 
We used perturbative continuous unitary transformation method to study 
the 1D extended ionic Hubbard model in the band insulating regime. 
EIHM in addition to the usual ionic Hubbard model includes 
a n.n. neighbor Coulomb repulsion, $V$. 
We treated the ionic potential as unperturbed part and studied the effect of 
hopping and Coulomb interaction terms, perturbatively. 
The convergence range of the perturbative results were extended by
using a Pad\'e scheme to fit the results and extrapolating them to
larger values of interaction parameters $U,V$ which enables us to
approach the limit where band gap vanishes. Effective Hamiltonian 
was obtained and diagonalized in zero-, one-particle and particle-hole sectors. 
Zero-particle sector gives us the ground state energy which upon using the
Hellman-Feynman theorem lead to a perturbative expression for the ionicity. 
Ionicity in the absence of $U,V$ terms at zero temperature has its maximum value
of $1$, which is a characteristic of band insulators. Ionicity decreases by increasing
$U$, while by increasing $V$, it tends to increase. Therefore the on-site correlations
$U$ are useful to remedy the valence discrepancy in transition metal oxides 
such as BaTiO$_3$, KNbO$_3$, and KTaO$_3$. 

In one-quasi-particle sector we calculated the electron and hole dispersions
which was achieved by adding one electron or a hole to the system. Study of the
charge gap at $V=0$ indicated that the band gap tends to zero for a critical value
of $U$ which is in agreement with state of the art DMRG results~\cite{Manmana,Ye}.

In the particle-hole sector, the total spin of particle-hole pair is a good quantum
number which helps to classify particle-hole excitations in singlet and triplet channels.
The particle-hole continuum is characterized by the spectrum of free particle-hole pairs.
Turning on the interaction gives rise to split-off (bound or anti-bound) states in
the triplet and singlet PH channels. There are in general three/two singlet and two triplet
branches in the reduced Brillouin zone ($-\pi/2\le k < \pi/2$). Given that the bound
states in each total spin channel are degenerate at $k=\pm \pi/2$, the two branches
can be thought of as a single branch in the larger BZ (corresponding to the $\Delta=0$ limit).
The so obtained triplet and singlet correspond to a pole in the appropriate susceptibility 
when the problem is viewed from itinerant limit. In this sense we can interpret the
triplet and singlet split-off states as collective modes. 
We also discussed the coherence length of these collective states. 
In general when the binding (anti-binding) energy of the collective states
which is defined as the energy difference between the energy eigen-value corresponding
to the collective state and the lower (upper) bound of the PHC decreases the
coherence length which is a measure of PH bound- (anti-bound-) state increases.
At a critical value of center of mass momentum $K$ where the collective mode branch
enters the continuum of free particle-hole pairs the coherence length becomes
infinitely large. 
Within PCUT when the $U$ term is zero, for finite $V$ the singlet and triplet
collective modes are almost degenerate. The major effect of n.n. Coulomb repulsion $V$ is to render 
the singlet branches into binding collective states by pushing them below the PHC, while the major
role of $U$ is to give rise to remarkable singlet-triplet splitting~\cite{BaskaranJafari}.
Both singlet and triplet states which will be bound states in presence of $V$ can be
interpreted as the excitonic states of the band insulator under study. Excitonic
states are well-known for their optical signatures. Therefore the PCUT approach seems
to be promising for investigation of excitonic effects in the optical spectra\cite{Schmidt1,Schmidt4}. 
Within the PCUT formulation it is also possible to discuss the propagation of two-electron
and two-hole states which touches upon the problem of superconductivity in strongly
correlated electron systems.

\section {Acknowledgement}
S.A.J. thanks G. S. Uhrig for critical reading of the manuscript and
many useful comments and suggestions. 
This research was supported by the Vice Chancellor for Research Affairs of 
the Isfahan University of Technology (IUT). S.A.J was supported by the 
National Elite Foundation (NEF) of Iran.

\begin{widetext}
\section*{Appendix}
Ground state energy and hopping terms up to order $9$ in the exponent summation 
of $t$, $U$, and $V$.
\bearr
\epsilon_0 &=& -\frac{1}{2}+\frac{1}{2}  U-\frac{1}{8} {t}^{2}
+\frac{3}{16} {t}^{2}V-\frac{1}{8} {t}^{2}U-{\frac {9}{32}}
 {t}^{2}{V}^{2}+\frac{3}{8} {t}^{2}UV-\frac{1}{8} {t}^{2}{U}^{2}+{\frac {3}{128}} 
{t}^{4}+{\frac {27}{64}} {t}^{2}{V}^{3}-{\frac {27}{32}} {t}^{2}U{V}
^{2}  \nn \\ && + 
{\frac {9}{16}} {t}^{2}{U}^{2}V-\frac{1}{8} {t}^{2}{U}^{3}-\frac{1}{8} {t}^{4}
V+{\frac {11}{128}} {t}^{4}U-{\frac {81}{128}} {t}^{2}{V}^{4}+{
\frac {27}{16}} {t}^{2}U{V}^{3}-{\frac {27}{16}} {t}^{2}{U}^{2}{V}^{
2}+\frac{3}{4} {t}^{2}{U}^{3}V-\frac{1}{8} {t}^{2}{U}^{4}   \nn \\ && 
+{\frac {109}{256}} {t}^{4
}{V}^{2}-{\frac {37}{64}} {t}^{4}UV+{\frac {49}{256}} {t}^{4}{U}^{2}
-{\frac {5}{512}} {t}^{6}+{\frac {243}{256}} {t}^{2}{V}^{5}-{\frac {
405}{128}} {t}^{2}U{V}^{4}+{\frac {135}{32}} {t}^{2}{U}^{2}{V}^{3}-{
\frac {45}{16}} {t}^{2}{U}^{3}{V}^{2}+{\frac {15}{16}} {t}^{2}{U}^{4
}V    \nn \\ &&
-\frac{1}{8} {t}^{2}{U}^{5}-{\frac {2407}{2048}} {t}^{4}{V}^{3}+{\frac {19
}{8}} {t}^{4}U{V}^{2}-{\frac {201}{128}} {t}^{4}{U}^{2}V+{\frac {175
}{512}} {t}^{4}{U}^{3}+{\frac {97}{1024}} {t}^{6}V-{\frac {17}{256}}
 {t}^{6}U-{\frac {729}{512}} {t}^{2}{V}^{6}+{\frac {729}{128}} {t}^
{2}U{V}^{5}   \nn \\ &&
-{\frac {1215}{128}} {t}^{2}{U}^{2}{V}^{4}+{\frac {135}{16
}} {t}^{2}{U}^{3}{V}^{3}-{\frac {135}{32}} {t}^{2}{U}^{4}{V}^{2}+{
\frac {9}{8}} {t}^{2}{U}^{5}V-\frac{1}{8} {t}^{2}{U}^{6}+{\frac {23459}{8192
}} {t}^{4}{V}^{4}-{\frac {7861}{1024}} {t}^{4}U{V}^{3}+{\frac {7791}
{1024}} {t}^{4}{U}^{2}{V}^{2}   \nn \\ &&
-{\frac {1699}{512}} {t}^{4}{U}^{3}V+{
\frac {551}{1024}} {t}^{4}{U}^{4}-{\frac {267}{512}} {t}^{6}{V}^{2}+
{\frac {2939}{4096}} {t}^{6}UV-{\frac {245}{1024}} {t}^{6}{U}^{2}+{
\frac {175}{32768}} {t}^{8}+{\frac {2187}{1024}} {t}^{2}{V}^{7}-{
\frac {5103}{512}} {t}^{2}U{V}^{6}  \nn \\ &&
+{\frac {5103}{256}} {t}^{2}{U}^{2
}{V}^{5}-{\frac {2835}{128}} {t}^{2}{U}^{3}{V}^{4}+{\frac {945}{64}}
 {t}^{2}{U}^{4}{V}^{3}-{\frac {189}{32}} {t}^{2}{U}^{5}{V}^{2}+{
\frac {21}{16}} {t}^{2}{U}^{6}V-\frac{1}{8} {t}^{2}{U}^{7}-{\frac {210403}{
32768}} {t}^{4}{V}^{5}   \nn \\ &&
+{\frac {175677}{8192}} {t}^{4}U{V}^{4}-{
\frac {58025}{2048}} {t}^{4}{U}^{2}{V}^{3}+{\frac {38025}{2048}} {t}
^{4}{U}^{3}{V}^{2}-{\frac {3093}{512}} {t}^{4}{U}^{4}V+{\frac {1599}{
2048}} {t}^{4}{U}^{5}+{\frac {70447}{32768}} {t}^{6}{V}^{3}-{\frac {
35913}{8192}} {t}^{6}U{V}^{2}  \nn \\ && 
+{\frac {23841}{8192}} {t}^{6}{U}^{2}V-
{\frac {81}{128}} {t}^{6}{U}^{3}-{\frac {1257}{16384}} {t}^{8}V+{
\frac {1787}{32768}} {t}^{8}U
\label{epsilon0.eqn}
\eearr

\bearr
t_{0,e} &=& t_{0,h}+U+2V=\frac{1}{2}+2\,V+\frac{1}{8}\,{t}^{2}-\frac{3}{8}\,{t}^{2}V
+\frac{1}{4}\,{t}^{2}U+\frac{3}{4}\,{t}^{2}{V}^{2}-
{t}^{2}UV+\frac{1}{4}\,{t}^{2}{U}^{2}-{\frac {3}{128}}\,{t}^{4}-{\frac {21}{16
}}\,{t}^{2}{V}^{3} \nn \\ &&
+{\frac {21}{8}}\,{t}^{2}U{V}^{2}-\frac{3}{2}\,{t}^{2}{U}^{2
}V+\frac{1}{4}\,{t}^{2}{U}^{3}+\frac{1}{4}\,{t}^{4}V-{\frac {11}{64}}\,{t}^{4}U+{
\frac {69}{32}}\,{t}^{2}{V}^{4}-{\frac {23}{4}}\,{t}^{2}U{V}^{3}+{
\frac {21}{4}}\,{t}^{2}{U}^{2}{V}^{2}-2\,{t}^{2}{U}^{3}V  \nn \\ &&
+\frac{1}{4}\,{t}^{2}{
U}^{4}-{\frac {37}{32}}\,{t}^{4}{V}^{2}+{\frac {99}{64}}\,{t}^{4}UV-{
\frac {53}{128}}\,{t}^{4}{U}^{2}+{\frac {5}{512}}\,{t}^{6}-{\frac {219
}{64}}\,{t}^{2}{V}^{5}+{\frac {365}{32}}\,{t}^{2}U{V}^{4}-{\frac {115}
{8}}\,{t}^{2}{U}^{2}{V}^{3}+{\frac {35}{4}}\,{t}^{2}{U}^{3}{V}^{2} \nn \\ &&
-\frac{5}{2}
\,{t}^{2}{U}^{4}V+\frac{1}{4}\,{t}^{2}{U}^{5}+{\frac {7769}{2048}}\,{t}^{4}{V}
^{3}-{\frac {1927}{256}}\,{t}^{4}U{V}^{2}+{\frac {563}{128}}\,{t}^{4}{
U}^{2}V-{\frac {393}{512}}\,{t}^{4}{U}^{3}-{\frac {97}{512}}\,{t}^{6}V
+{\frac {17}{128}}\,{t}^{6}U+{\frac {681}{128}}\,{t}^{2}{V}^{6} \nn \\ &&
-{
\frac {681}{32}}\,{t}^{2}U{V}^{5}+{\frac {1095}{32}}\,{t}^{2}{U}^{2}{V
}^{4}-{\frac {115}{4}}\,{t}^{2}{U}^{3}{V}^{3}+{\frac {105}{8}}\,{t}^{2
}{U}^{4}{V}^{2}-3\,{t}^{2}{U}^{5}V+\frac{1}{4}\,{t}^{2}{U}^{6}-{\frac {84403}{
8192}}\,{t}^{4}{V}^{4}+{\frac {55553}{2048}}\,{t}^{4}U{V}^{3} \nn \\ &&
-{\frac {
50827}{2048}}\,{t}^{4}{U}^{2}{V}^{2}+\frac{19}{2}\,{t}^{4}{U}^{3}V-{\frac {
1265}{1024}}\,{t}^{4}{U}^{4}+{\frac {5843}{4096}}\,{t}^{6}{V}^{2}-{
\frac {3941}{2048}}\,{t}^{6}UV+{\frac {275}{512}}\,{t}^{6}{U}^{2}-{
\frac {175}{32768}}\,{t}^{8}-{\frac {2091}{256}}\,{t}^{2}{V}^{7} \nn \\ &&
+{
\frac {4879}{128}}\,{t}^{2}U{V}^{6}-{\frac {4767}{64}}\,{t}^{2}{U}^{2}
{V}^{5}+{\frac {2555}{32}}\,{t}^{2}{U}^{3}{V}^{4}-{\frac {805}{16}}\,{
t}^{2}{U}^{4}{V}^{3}+{\frac {147}{8}}\,{t}^{2}{U}^{5}{V}^{2}-\frac{7}{2}\,{t}^
{2}{U}^{6}V+\frac{1}{4}\,{t}^{2}{U}^{7} \nn \\ &&
+{\frac {814169}{32768}}\,{t}^{4}{V}^{5
}-{\frac {1337047}{16384}}\,{t}^{4}U{V}^{4}+{\frac {208933}{2048}}\,{t
}^{4}{U}^{2}{V}^{3}-{\frac {251523}{4096}}\,{t}^{4}{U}^{3}{V}^{2}+{
\frac {35873}{2048}}\,{t}^{4}{U}^{4}V-{\frac {3725}{2048}}\,{t}^{4}{U}
^{5} \nn \\ &&
-{\frac {231121}{32768}}\,{t}^{6}{V}^{3}+{\frac {115041}{8192}}\,{
t}^{6}U{V}^{2}-{\frac {68499}{8192}}\,{t}^{6}{U}^{2}V+{\frac {769}{512
}}\,{t}^{6}{U}^{3}+{\frac {1257}{8192}}\,{t}^{8}V-{\frac {1787}{16384}
}\,{t}^{8}U
\eearr

\bearr
t_2 &=& \Big(\frac{1}{16}-\frac{1}{16}\,V+\frac{1}{16}\,U+\frac{1}{16}\,{V}^{2}
-\frac{1}{8}\,UV+\frac{1}{16}\,{U}^{2}-{
\frac {1}{64}}\,{t}^{2}-\frac{1}{16}\,{V}^{3}+\frac{3}{16}\,U{V}^{2}
-\frac{3}{16}\,{U}^{2}V+\frac{1}{16}\,{U}^{3}  
 \nn \\ &&
+{\frac {9}{128}}\,{t}^{2}V-\frac{1}{16}\,{t}^{2}U+\frac{1}{16}\,{V}^{4}
-\frac{1}{4}\,U{V}^{3}+\frac{3}{8}\,{U}^{2}{V}^{2}-\frac{1}{4}\,{U}^{3}V+\frac{1}{16}\,{U}^{4}-{\frac {99
}{512}}\,{t}^{2}{V}^{2}+{\frac {165}{512}}\,{t}^{2}UV-{\frac {9}{64}}
\,{t}^{2}{U}^{2}  \nn \\ &&
+{\frac {15}{2048}}\,{t}^{4}-\frac{1}{16}\,{V}^{5}+{\frac {5}{
16}}\,U{V}^{4}-\frac{5}{8}\,{U}^{2}{V}^{3}+\frac{5}{8}\,{U}^{3}{V}^{2}-{\frac {5}{16}}
\,{U}^{4}V+\frac{1}{16}\,{U}^{5}+{\frac {879}{2048}}\,{t}^{2}{V}^{3}-{\frac {
2113}{2048}}\,{t}^{2}U{V}^{2}   \nn \\ &&
+{\frac {889}{1024}}\,{t}^{2}{U}^{2}V-\frac{1}{4}
\,{t}^{2}{U}^{3}-{\frac {135}{2048}}\,{t}^{4}V+{\frac {117}{2048}}\,{t
}^{4}U+\frac{1}{16}\,{V}^{6}-\frac{3}{8}\,U{V}^{5}+{\frac {15}{16}}\,{U}^{2}{V}^{4}
-\frac{5}{4}\,{U}^{3}{V}^{3}+{\frac {15}{16}}\,{U}^{4}{V}^{2}    \nn \\ &&
-\frac{3}{8}\,{U}^{5}V+\frac{1}{16}
\,{U}^{6}-{\frac {6965}{8192}}\,{t}^{2}{V}^{4}+{\frac {21737}{8192}}\,
{t}^{2}U{V}^{3}-{\frac {13}{4}}\,{t}^{2}{U}^{2}{V}^{2}+{\frac {3721}{
2048}}\,{t}^{2}{U}^{3}V-{\frac {25}{64}}\,{t}^{2}{U}^{4}+{\frac {20605
}{65536}}\,{t}^{4}{V}^{2}    \nn \\ &&
-{\frac {1037}{2048}}\,{t}^{4}UV+{\frac {3449
}{16384}}\,{t}^{4}{U}^{2}-{\frac {35}{8192}}\,{t}^{6}-\frac{1}{16}\,{V}^{7}+{
\frac {7}{16}}\,U{V}^{6}-{\frac {21}{16}}\,{U}^{2}{V}^{5}+{\frac {35}{
16}}\,{U}^{3}{V}^{4}-{\frac {35}{16}}\,{U}^{4}{V}^{3}+{\frac {21}{16}}
\,{U}^{5}{V}^{2}    \nn \\ &&
-{\frac {7}{16}}\,{U}^{6}V+\frac{1}{16}\,{U}^{7}+{\frac {51655
}{32768}}\,{t}^{2}{V}^{5}-{\frac {197589}{32768}}\,{t}^{2}U{V}^{4}+{
\frac {157273}{16384}}\,{t}^{2}{U}^{2}{V}^{3}-{\frac {64179}{8192}}\,{
t}^{2}{U}^{3}{V}^{2}+{\frac {13425}{4096}}\,{t}^{2}{U}^{4}V      \nn \\ &&
-{\frac {9}
{16}}\,{t}^{2}{U}^{5}-{\frac {72489}{65536}}\,{t}^{4}{V}^{3}+{\frac {
41967}{16384}}\,{t}^{4}U{V}^{2}-{\frac {33523}{16384}}\,{t}^{4}{U}^{2}
V+{\frac {4557}{8192}}\,{t}^{4}{U}^{3}+{\frac {493}{8192}}\,{t}^{6}V-{
\frac {53}{1024}}\,{t}^{6}U \Big) {t}^{2}
\eearr

\bearr
t_4 &=& \Big( -{\frac {1}{256}}+{\frac {1}{64}}\,V-{\frac {3}{256}}\,
U-{\frac {85}{2048}}\,{V}^{2}+\frac{1}{16}\,UV-{\frac {3}{128}}\,{U}^{2}+{
\frac {3}{1024}}\,{t}^{2}+{\frac {47}{512}}\,{V}^{3}-{\frac {425}{2048
}}\,U{V}^{2}+{\frac {5}{32}}\,{U}^{2}V     \nn \\ &&
-{\frac {5}{128}}\,{U}^{3} 
-{\frac {3}{128}}\,{t}^{2}V+{\frac {9}{512}}\,{t}^{2}U-{\frac {1499}{
8192}}\,{V}^{4}+{\frac {141}{256}}\,U{V}^{3}-{\frac {1275}{2048}}\,{U}
^{2}{V}^{2}+{\frac {5}{16}}\,{U}^{3}V-{\frac {15}{256}}\,{U}^{4}+{
\frac {881}{8192}}\,{t}^{2}{V}^{2}     \nn \\ &&
-{\frac {329}{2048}}\,{t}^{2}UV+{
\frac {15}{256}}\,{t}^{2}{U}^{2}-{\frac {35}{16384}}\,{t}^{4}+{\frac {
2797}{8192}}\,{V}^{5}-{\frac {10493}{8192}}\,U{V}^{4}+{\frac {987}{512
}}\,{U}^{2}{V}^{3}-{\frac {2975}{2048}}\,{U}^{3}{V}^{2}+{\frac {35}{64
}}\,{U}^{4}V     \nn \\ &&
-{\frac {21}{256}}\,{U}^{5} 
-{\frac {24475}{65536}}\,{t}^{2
}{V}^{3}+{\frac {13637}{16384}}\,{t}^{2}U{V}^{2}-{\frac {1247}{2048}}
\,{t}^{2}{U}^{2}V+{\frac {75}{512}}\,{t}^{2}{U}^{3}+{\frac {55}{2048}}
\,{t}^{4}V-{\frac {83}{4096}}\,{t}^{4}U \Big) {t}^{4} 
\eearr

\bearr
t_6 &=& \Big( {\frac {1}{2048}}-{\frac {7}{2048}}\,V+{\frac {5}{2048}}\,U+{
\frac {951}{65536}}\,{V}^{2}-{\frac {21}{1024}}\,UV+{\frac {15}{2048}}
\,{U}^{2}-{\frac {5}{8192}}\,{t}^{2}-{\frac {6345}{131072}}\,{V}^{3}+{
\frac {6657}{65536}}\,U{V}^{2}    \nn \\ &&
-{\frac {147}{2048}}\,{U}^{2}V+{\frac {
35}{2048}}\,{U}^{3}+{\frac {111}{16384}}\,{t}^{2}V-{\frac {5}{1024}}\,
{t}^{2}U \Big) {t}^{6}
\eearr

\bearr
t_8= \Big( -{\frac {5}{65536}}+{\frac {25}{32768}}\,V-{\frac {35}{65536}}
\,U \Big) {t}^{8}
\eearr

Electron-hole energy $E(K)$ up to order $8$ at $K=\frac{\pi}{2}$ valid for $t^8\ll V$.
\bearr
E_{S1}(\frac{\pi}{2}) &=& E_{S2}(\frac{\pi}{2})=
1+ \frac{3}{2}\,V-U+\frac{1}{4}\,{t}^{2}-\frac{5}{8}\,{t}^{2}V+\frac{5}{8}\,{t}^{2}U+{\frac {17}{16}}\,
{t}^{2}{V}^{2}-{\frac {13}{8}}\,{t}^{2}UV+\frac{3}{8}\,{t}^{2}{U}^{2}  
-{\frac {3}{64}}\,{t}^{4} -{\frac {53}{32}}\,{t}^{2}{V}^{3}     \nn \\ &&
+{\frac {111}{32}}\,{t}^{2}U{V}^{2}-{\frac {33}{16}}\,{t}^{2}{U}^{2}V  
+\frac{5}{8}\,{t}^{2}{U}^{3}+{
\frac {217}{512}}\,{t}^{4}V-{\frac {27}{64}}\,{t}^{4}U+{\frac {161}{64
}}\,{t}^{2}{V}^{4}    
-{\frac {109}{16}}\,{t}^{2}U{V}^{3}+{\frac {105}{16}}\,{t}^{2}{U}^{2}{V}^{2}    \nn \\ &&
-{\frac {13}{4}}\,{t}^{2}{U}^{3}V+\frac{3}{8}\,{t}^{2}
{U}^{4}-{\frac {3357}{2048}}\,{t}^{4}{V}^{2}+{\frac {319}{128}}\,{t}^{
4}UV-{\frac {81}{128}}\,{t}^{4}{U}^{2}+{\frac {5}{256}}\,{t}^{6}  
-{\frac {485}{128}}\,{t}^{2}{V}^{5}+{\frac {1625}{128}}\,{t}^{2}U{V}^{4}      \nn \\ &&
-{\frac {535}{32}}\,{t}^{2}{U}^{2}{V}^{3}+{\frac {185}{16}}\,{t}^{2}{U
}^{3}{V}^{2}-{\frac {55}{16}}\,{t}^{2}{U}^{4}V+\frac{5}{8}\,{t}^{2}{U}^{5}+{
\frac {38801}{8192}}\,{t}^{4}{V}^{3}-{\frac {10115}{1024}}\,{t}^{4}U{V
}^{2}+{\frac {3067}{512}}\,{t}^{4}{U}^{2}V       \nn \\ &&
-{\frac {201}{128}}\,{t}^{4}
{U}^{3}-{\frac {667}{2048}}\,{t}^{6}V+{\frac {655}{2048}}\,{t}^{6}U-{
\frac {1}{4096}}\,{\frac {{t}^{8}}{V}}+{\frac {1457}{256}}\,{t}^{2}{V}
^{6}-{\frac {2919}{128}}\,{t}^{2}U{V}^{5}+{\frac {4845}{128}}\,{t}^{2}
{U}^{2}{V}^{4}-{\frac {545}{16}}\,{t}^{2}{U}^{3}{V}^{3}       \nn \\ &&
+{\frac {525}{32}}\,{t}^{2}{U}^{4}{V}^{2}-{\frac {39}{8}}\,{t}^{2}{U}^{5}V+\frac{3}{8}\,{t}^
{2}{U}^{6}-{\frac {386537}{32768}}\,{t}^{4}{V}^{4}+{\frac {65089}{2048
}}\,{t}^{4}U{V}^{3}-{\frac {31179}{1024}}\,{t}^{4}{U}^{2}{V}^{2}+{
\frac {7177}{512}}\,{t}^{4}{U}^{3}V       \nn \\ &&
-{\frac {477}{256}}\,{t}^{4}{U}^{4}
+{\frac {16737}{8192}}\,{t}^{6}{V}^{2}-{\frac {12677}{4096}}\,{t}^{6}U
V+{\frac {1701}{2048}}\,{t}^{6}{U}^{2}-{\frac {147}{16384}}\,{t}^{8}-{
\frac {7}{4096}}\,{\frac {{t}^{8}U}{V}}
\eearr

\bearr
E_{T1}(\frac{\pi}{2}) &=& E_{T2}(\frac{\pi}{2})=
1+\frac{3}{2}\,V-U+\frac{1}{4}\,{t}^{2}-\frac{5}{8}\,{t}^{2}V+\frac{3}{8}\,{t}^{2}U+{\frac {9}{8}}\,{t
}^{2}{V}^{2}-{\frac {11}{8}}\,{t}^{2}UV+\frac{3}{8}\,{t}^{2}{U}^{2}-{\frac {3}
{64}}\,{t}^{4}-{\frac {59}{32}}\,{t}^{2}{V}^{3}     \nn \\ &&
+{\frac {111}{32}}\,{t}
^{2}U{V}^{2}-{\frac {33}{16}}\,{t}^{2}{U}^{2}V+\frac{3}{8}\,{t}^{2}{U}^{3}+{
\frac {217}{512}}\,{t}^{4}V-{\frac {17}{64}}\,{t}^{4}U+{\frac {93}{32}
}\,{t}^{2}{V}^{4}-{\frac {119}{16}}\,{t}^{2}U{V}^{3}+{\frac {111}{16}}
\,{t}^{2}{U}^{2}{V}^{2}     \nn \\ &&
-\frac{11}{4}\,{t}^{2}{U}^{3}V+\frac{3}{8}\,{t}^{2}{U}^{4}-{
\frac {3597}{2048}}\,{t}^{4}{V}^{2}+{\frac {283}{128}}\,{t}^{4}UV-{
\frac {81}{128}}\,{t}^{4}{U}^{2}+{\frac {5}{256}}\,{t}^{6}-{\frac {575
}{128}}\,{t}^{2}{V}^{5}+{\frac {1865}{128}}\,{t}^{2}U{V}^{4}     \nn \\ &&
-{\frac {595}{32}}\,{t}^{2}{U}^{2}{V}^{3}+{\frac {185}{16}}\,{t}^{2}{U}^{3}{V}^
{2}-{\frac {55}{16}}\,{t}^{2}{U}^{4}V+\frac{3}{8}\,{t}^{2}{U}^{5}+{\frac {
43969}{8192}}\,{t}^{4}{V}^{3}-{\frac {10487}{1024}}\,{t}^{4}U{V}^{2}+{
\frac {3163}{512}}\,{t}^{4}{U}^{2}V     \nn \\ &&
-{\frac {149}{128}}\,{t}^{4}{U}^{3}
-{\frac {667}{2048}}\,{t}^{6}V+{\frac {433}{2048}}\,{t}^{6}U-{\frac {1
}{4096}}\,{\frac {{t}^{8}}{V}}+{\frac {879}{128}}\,{t}^{2}{V}^{6}-{
\frac {3453}{128}}\,{t}^{2}U{V}^{5}+{\frac {5595}{128}}\,{t}^{2}{U}^{2
}{V}^{4}-{\frac {595}{16}}\,{t}^{2}{U}^{3}{V}^{3}     \nn \\ &&
+{\frac {555}{32}}\,{
t}^{2}{U}^{4}{V}^{2}-{\frac {33}{8}}\,{t}^{2}{U}^{5}V+\frac{3}{8}\,{t}^{2}{U}^
{6}-{\frac {456009}{32768}}\,{t}^{4}{V}^{4}+{\frac {73161}{2048}}\,{t}
^{4}U{V}^{3}-{\frac {34055}{1024}}\,{t}^{4}{U}^{2}{V}^{2}+{\frac {6777
}{512}}\,{t}^{4}{U}^{3}V     \nn \\ &&
-{\frac {477}{256}}\,{t}^{4}{U}^{4}+{\frac {
18031}{8192}}\,{t}^{6}{V}^{2}-{\frac {11619}{4096}}\,{t}^{6}UV+{\frac 
{1725}{2048}}\,{t}^{6}{U}^{2}-{\frac {147}{16384}}\,{t}^{8}-{\frac {5}
{4096}}\,{\frac {{t}^{8}U}{V}}
\eearr
\end{widetext}

\end{document}